\RequirePackage{lineno} 
\documentclass[aps,prc,superscriptaddress,showpacs,nofootinbib,floatfix,twocolumn]{revtex4}
\usepackage{graphicx} 
\usepackage{float} 
\usepackage{amssymb}
\usepackage{url}
\usepackage{hyperref}
\newcommand{\pT} {\ensuremath{p_{\mathrm{T}}}}
 
\begin{document} % do not change
\title{Azimuthal anisotropy in a jet absorption model with fluctuating initial geometry in heavy ion collisions}
\newcommand{\sunysb}{Department of Chemistry, Stony Brook University, Stony Brook, NY 11794, USA}
\newcommand{\bnl}{Physics Department, Brookhaven National Laboratory, Upton, NY 11796, USA}
\author{Jiangyong Jia}
\affiliation{\sunysb}\affiliation{\bnl}
\date{\today}
\begin{abstract}
\linenumbers 
The azimuthal anisotropy due to path-length dependent jet energy loss is studied in a simple jet absorption model that include event by event fluctuating Glauber geometry. Significant anisotropy coefficients $v_n$ are observed for $n=1$,2 and 3, but they are very small for $n>3$. These coefficients are expected to result in a ``ridge'' for correlations between two independently produced jets. The correlations between the orientation of the $n^{\mathrm{th}}$-order anisotropy induced by jet absorption ($\Phi_n^{\mathrm{QP}}$) and the $n^{\mathrm{th}}$-order participant plane ($\Phi_n^{\mathrm {PP}}$) responsible for harmonic flow are studied. Tight correlations are observed for $n=2$ in mid-central collisions, but they weaken significantly for $n\neq2$. The correlations are positive for $n\leq3$, but become negative in central collisions for $n>3$. The dispersion between $\Phi_n^{\mathrm{QP}}$ and $\Phi_n^{\mathrm{PP}}$ is expect to break the factorization of the Fourier coefficients from two-particle correlation $v_{n,n}$ into the single particle $v_n$, and has important implications for the high-$\pT$ ridge phenomena. 
\end{abstract}
 \maketitle \linenumbers

\section{Introduction}

Recently, a lot of attentions are focused on the study of the azimuthal anisotropy of the particle production in heavy ion collisions at the Relativistic Heavy Ion Collider (RHIC) and the Large Hadron Collider (LHC). This anisotropy is usually expanded into a Fourier series:
\begin{eqnarray}
\label{eq:1}
\frac{dN}{d\phi}&\propto& 1+2\sum_{n=1}^{\infty} v_{n} \cos n(\phi-\Phi_n)
\end{eqnarray}
with $v_n$ and $\Phi_n$ represent the magnitude and direction of $n^{\mathrm{\mathrm{th}}}$-order anisotropy, respectively. At low $\pT$, $v_n$ is driven by the anisotropic pressure gradient associated with the initial spatial asymmetries, with more particles emitted in the direction of largest gradients~\cite{Ollitrault:1992bk}. Asymmetries giving rise to non-zero $v_n$ are associated with either average shape (for $n=2$) or shapes arising from spatial fluctuations of the participating nucleons~\cite{Alver:2010gr,Alver:2010dn,Staig:2010pn,Teaney:2010vd}. They can be characterized by a set of multi-pole components (also known as ``eccentricities'') at different angular scale, calculated from the participating nucleons at $(r,\phi)$~\cite{Alver:2010dn}:
\begin{eqnarray}
\label{eq:ena}
\epsilon_n = \frac{\sqrt{\langle r^2\cos n\phi\rangle^2+\langle r^2\sin n\phi\rangle^2}}{\langle r^2\rangle}.
\end{eqnarray}
The orientations of the minor axis for each moment $n$, also known as the participant plane (PP)  are given by
\begin{eqnarray}
\label{eq:enb}
\Phi_n^{\mathrm{PP}}=\frac{atan2(\langle r^2\sin n\phi\rangle,\langle r^2\cos n\phi\rangle)}{n}+\frac{\pi}{n}
\end{eqnarray}
When fluctuations are small and linearized hydrodynamics is applicable, each moment of the flow $v_n$ is expected to be independently driven by $\epsilon_n$ along $\Phi_n^{\mathrm{PP}}=\Phi_n$~\cite{Alver:2010dn}. This may not be true when the fluctuations are large, as the non-linear effects may lead to significant mixing between harmonic flow of different order~\cite{Qiu:2011iv}. In this paper, they are assumed to be the same to facilitate the study of the correlations between $\Phi_n$ of different physics origins. 

At high $\pT$ ($\pT\gtrsim10$) GeV, the $v_n$ is understood to be driven by the path-length dependent energy loss of jets traversing the medium, with more particles emitted along the direction of shortest path-length (or direction of smallest jet attenuation)~\cite{Gyulassy:2000gk,Adare:2010sp}. In contrast to flow which is sensitive to the global geometry manifested through the global evolution of the created matter, this anisotropy is more sensitive to the local density experienced by the jets as they traverse the matter. Nevertheless, since both flow and jet quenching are influenced by the same geometry, the directions of largest pressure gradient for flow and direction of smallest jet attenuation are strongly correlated. In fact, they are often implicitly assumed to be the same in many theoretical calculations~\cite{Bass:2008rv,Jia:2010ee,Jia:2011pi,Renk:2011qi,Betz:2011tu,Zhang:2012mi}. An explicit study of the correlation between the two directions can help clarifying this assumption.

In this paper, we estimate the high-$\pT$ anisotropy coefficients $v_n$ and their associated directions $\Psi_n^{\mathrm{QP}}$ (QP stands for ``quenching plane'') using a simple jet absorption model with event by event fluctuating Glauber geometry. We study the correlations between $\Psi_n^{\mathrm{PP}}$ and $\Psi_n^{\mathrm{QP}}$, and discuss implications of these correlations for the interpretation of the ``ridge'' phenomena in two-particle correlations (2PC).

\section{Model}
\begin{figure*}[!t]
\includegraphics[width= 1\linewidth]{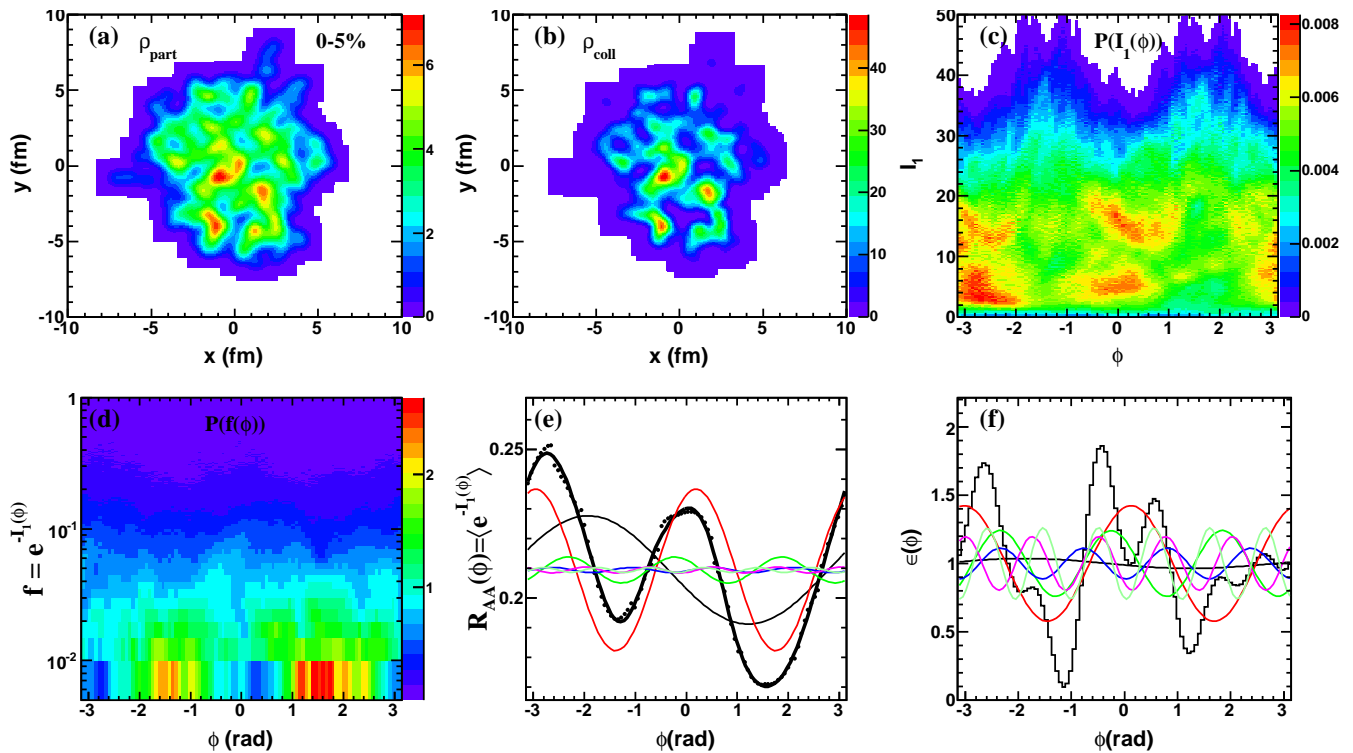}
\caption{(Color online) The complete set of output obtained in the jet absorption model for one event in 0-5\% centrality interval: (a) the participant density profile ($\rho_{p}$) ; (b) the collision density profile ($\rho_{c}$), (c) the probability distribution of the path-length integral $I_1$, (d) the probability distribution of jet surviving the exponential attenuation, (e) the distribution of survival rate as function of azimuth angle, (f) the initial spatial asymmetry of the participants calculated via Eq.~\ref{eq:5}. The original impact parameter of the event is aligned along the $x$-axis.}
\label{fig:1}
\end{figure*}

We use a simple jet absorption/Glauber model of Ref.~\cite{Drees:2003zh} to calculate $v_n$ and $\Psi_n^{\mathrm{QP}}$. This model has been used previously to study the centrality and path-length dependence of single particle suppression $R_{\mathrm {AA}}$, dihadron suppression $I_{\mathrm {AA}}$ and $v_2$. Back-to-back jet pairs are generated according to the binary collision density profile ($\rho_{c}$) in the transverse ($xy$) plane with random orientation. They are then propagated through the medium, whose density is given by the participant density profile ($\rho_{p}$). Both profiles are generated with a Monte Carlo Glauber model with event by event fluctuation of positions of nucleons in Au ions~\cite{Alver:2008aq}. The nucleons are sampled from a Woods-Saxon distribution with a radius of 6.38 fm and diffuseness of 0.535 fm, with a nucleon-nucleon cross-section of $\sigma_{\mathrm {nn}}=42$~mb.  In order to have smooth distributions for $\rho_{c}$ and $\rho_{p}$, the nucleons are assumed to have a Gaussian profile in transverse plane with a width of $r_{0}=0.4$~fm in $x$ and $y$ direction similar to Ref~\cite{Jia:2010ee}. The value of $r_0$ is varied from 0.2-0.4~fm, and the nucleon is also assumed to be a uniform disk with a radius of $\sqrt{\sigma_{\mathrm {nn}}/\pi}/2=0.58$~fm. However the final results are found to be insensitive to the details of the nucleon shape, except in peripheral collisions.

The jet quenching is implemented via exponential attenuation $f=e^{-\kappa I}$, where the matter integral $I$ is calculated as 
\begin{eqnarray}
\label{eq:3}
I_m &=& \int_{0}^\infty dl\frac{l^{m}}{l+l_0}
\rho{\left(\overrightarrow{\mathbf{r}}+\left(l+l_0\right)\widehat{\mathbf{v}}\right)} \\ &\approx& \int_{0}^\infty dl\hspace{1mm}l^{m-1}
\hspace{1mm}\rho{\left(\overrightarrow{\mathbf{r}}+l\widehat{\mathbf{v}}\right)},\;\;\;\;\; m=1,2.
\end{eqnarray}
for jet generated at $\overrightarrow{\mathbf{r}}=(x,y)$ and propagated along direction $\widehat{\mathbf{v}}$. They corresponds to $l^{m+1}$ dependence of absorption ($\propto l^{m}dl$) in a longitudinal expanding or 1+1D medium ($\propto 1/(l_0+l)$) with a thermalization time of $l_0=c\tau_0$. The $l_0$ is fixed to 0 by default, but we have checked the $v_n$ do not change much for $l_0<0.3$~fm~\cite{Jia:2010ee}. The two cases, $m=1$ and $m=2$, are motivated for the $l$ dependence expected for radiative and AdS/CFT energy loss in 1+1D medium~\cite{Chesler:2008uy,Marquet:2009eq}, respectively.

The absorption coefficient $\kappa$ controls the jet quenching strength and is the only parameter in this calculation. It is tuned to reproduce $R_{\rm AA}=\langle e^{-\kappa I_m}\rangle\sim0.19$ for 0-5\% $\pi^0$ data at RHIC after averaging over many Glauber events~\cite{Adare:2008qa}. This leads to a value of $\kappa=0.1473$~fm$^{-1}$ and 0.0968~fm$^{-2}$ for $m=1$ and 2, respectively. 

\section{Results}
Figure~\ref{fig:1} summarize the basic information obtained from this procedure for one typical Au-Au event in 0-5\% centrality interval. Panels (a) and (b) shows the density profile for $\rho_{p}$ and $\rho_{c}$, respectively. Panel (c) shows the normalized probability distribution of $I_1$: $P(I_1(\phi))$, which is obtained by calculating $I_1$ over all possible di-jet production point $\rho_{c}$ and jet propagation direction $\phi$. This distribution exhibit characteristic high density and low density regions in $(I_1,\phi)$ space, presumably reflecting spatial correlation between the $\rho_{c}$ and $\rho_{p}$ profiles. Panel (d) shows the normalized probability distribution of the attenuation $e^{-\kappa I_1}$. Panel (e) shows the $\langle e^{-\kappa I_1}\rangle$ averaged along the y-axis in Panel (d) as a function $\phi$, which is precisely the azimuthal angle dependent suppression $R_{\mathrm {AA}}(\phi)$. A clear anti-correlation can be seen between the peak magnitude of the $R_{\mathrm {AA}}(\phi)$ and breadth of the $I_1$ distribution in Panel (c). This distribution can also be obtained by randomly generating many di-jet pairs according the $\rho_{c}$ and propagating them through $\rho_{p}$ via Eq.~\ref{eq:4}. We expand it into a Fourier series:
\begin{eqnarray}
\label{eq:4}
R_{\mathrm {AA}}(\phi) = R_{\mathrm {AA}}^{0}(1+2\sum_{n=1}^{\infty} v_{n}^{\mathrm{QP}} \cos n(\phi-\Phi_n^{\mathrm{QP}}))\;,
\end{eqnarray}
where $R_{\mathrm {AA}}^0$ represents the average suppression, $v_{n}^{\mathrm{QP}}$ and $\Phi_n^{\mathrm{QP}}$ represent the magnitude and direction of $n^{\mathrm{th}}$-order harmonic of emission probability distribution, respectively. Similar studies of $R_{\mathrm {AA}}(\phi)$ were pursued before in Ref.~\cite{Renk:2011qi} for a pQCD energy loss in a event by event hydrodynamic underlying event. However it focused primarily on the influence of fluctuations on the event-averaged $R_{\mathrm {AA}}(\phi)$ distribution relative to the $2^{\mathrm{nd}}$-order event plane (EP).

Figure~\ref{fig:1} (f) shows a distribution calculated from $\epsilon_n$ and $\Phi_n^{\mathrm{QP}}$:
\begin{eqnarray}
\label{eq:5}
\epsilon(\phi) = 1+2\sum_{n=1}^{\infty} \epsilon_{n} \cos n(\phi-\Phi_n^{\mathrm{PP}})\;.
\end{eqnarray}
It visualizes the shape of the initial geometry that is transformed into the final momentum anisotropy via either flow or jet quenching. A good alignment is seen between $\Phi_n^{\mathrm{PP}}$ and $\Phi_n^{\mathrm{QP}}$ for $n\le3$. It also shows that the large $\epsilon_n$ for $n>3$ are strongly damped after jet absorption, leading to very small values of $v_n^{\mathrm{QP}}$ for $n>3$. 

\begin{figure}[!t]
\includegraphics[width= 1\linewidth]{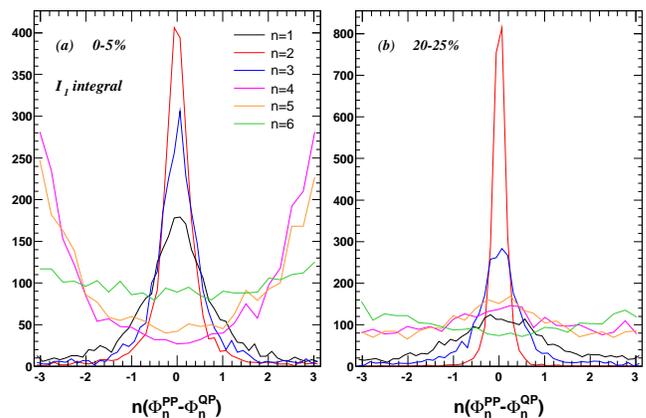}
\caption{(Color online) The correlation between participant plane $\Phi_n^{\mathrm{PP}}$ and quenching plane $\Phi_n^{\mathrm{QP}}$ for $n=1-6$ calculated for $I_1$ path-length dependence and for (a) 0-5\% and (b) 20-25\% centrality interval.}
\label{fig:2}
\end{figure}

The study shown in Fig.~\ref{fig:1} can be repeated for many events. We divide the simulation data into 5\% centrality intervals, each containing about 2500 events. Figure~\ref{fig:2} shows the distribution of $\Phi_n^{\mathrm{PP}}-\Phi_n^{\mathrm{QP}}$ for two centrality intervals. Strong positive correlations are obtained for $n=1$, 2 and 3, while the correlations are rather weak or even become negative for $n>3$~\footnote{$\Phi_n^{\mathrm{PP}}$ in Eq. is calculated with $r^2$ weighting. We have also repeated the study using $r^n$ weighting for $n>1$ and $r^3$ weighing for $n=1$~\cite{Teaney:2010vd}, but very little differences are seen.}.

In heavy ion collisions at RHIC and LHC, the $v_n$ is usually measured from particle distribution relative to $\Psi_n$ via Eq.~\ref{eq:1}~\cite{Poskanzer:1998yz}. However it has also been derived from the Fourier coefficients of two-particle correlation in relative azimuthal angle $\Delta\phi=\phi^{\mathrm{a}}-\phi^{\mathrm{b}}$~\cite{ATLAS}:
\begin{eqnarray}
\label{eq:6a}
 \frac{dN_{\mathrm{pairs}}}{d\Delta\phi} \propto 1+2\sum_{n=1}^{\infty}v_{n,n}(\pT^{\mathrm{a}},\pT^{\mathrm b}) \cos n\Delta\phi\;,
\end{eqnarray} 
with
\begin{eqnarray}
\label{eq:6b}
v_{n,n}(\pT^{\mathrm a},\pT^{\mathrm b}) = v_n(\pT^{\mathrm a})v_n(\pT^{\mathrm b})\;.
\end{eqnarray} 
The fact that the quenching plane and participant plane do not align exactly with each other implies that the $v_n$ measured relative to $\Psi_n^{\mathrm{PP}}$ is not the same as those contributing to the 2PC in Eq.~\ref{eq:6a}. In other words, it is possible that the $v_n$ obtained from single particle analysis is only a fraction of the true anisotropy resulting from jet quenching:
\begin{eqnarray}
\label{eq:7}
v_n = v_n^{\mathrm{QP}} \langle\cos n(\Phi_n^{\mathrm{PP}}-\Phi_n^{\mathrm{QP}})\rangle
\end{eqnarray}
Since what is measured in experiment is the event plane not the PP, it is important to check whether the event plane align with QP or not, for example in a hydrodynamic model calculation.
%Note that what is measured in experiment is the event plane not the PP, thus it is more relavent to check whether the event plane align with QP or not, for example in a hydrodynamic model calculation.

\begin{figure}[!t]
\includegraphics[width= 1\linewidth]{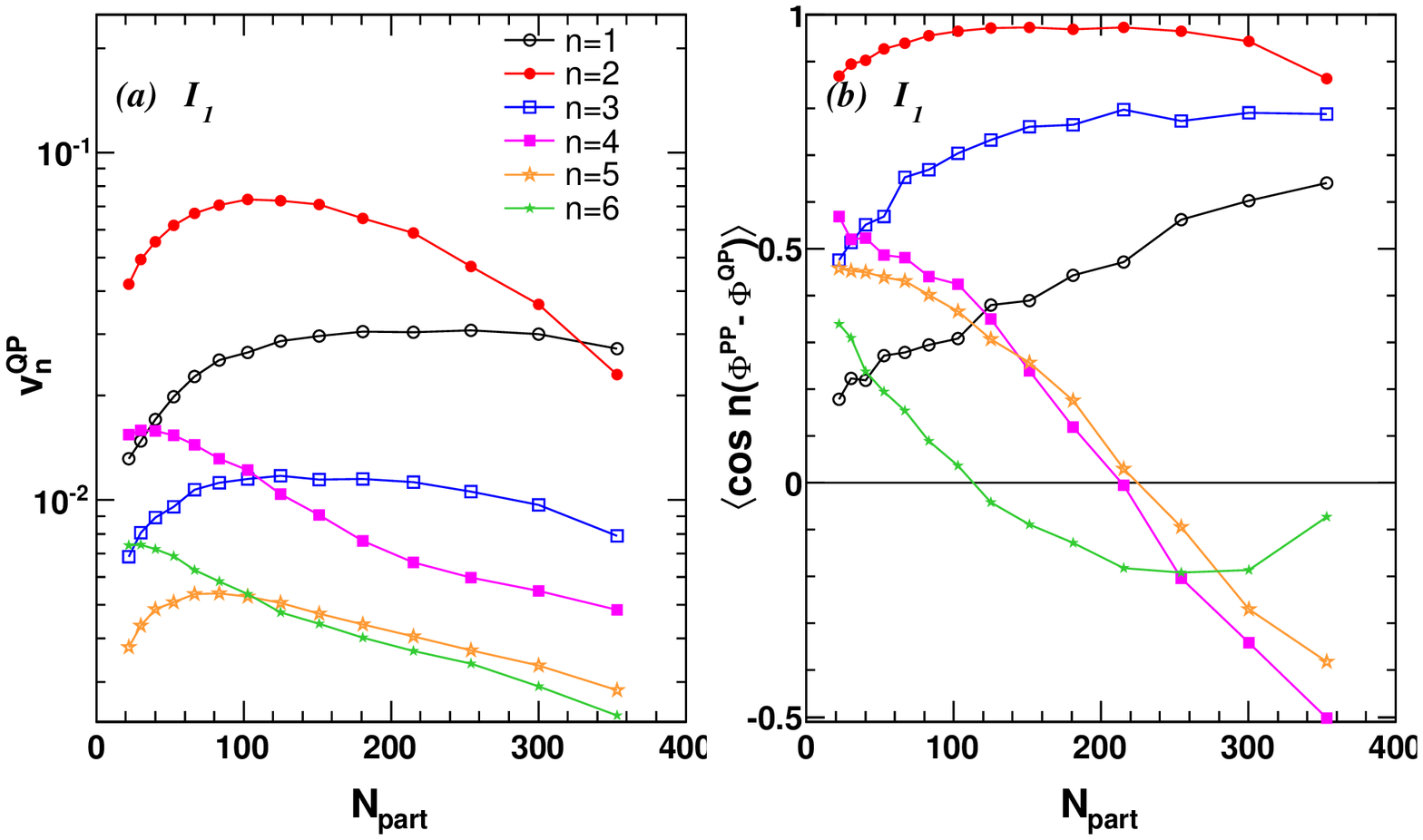}
\includegraphics[width= 1\linewidth]{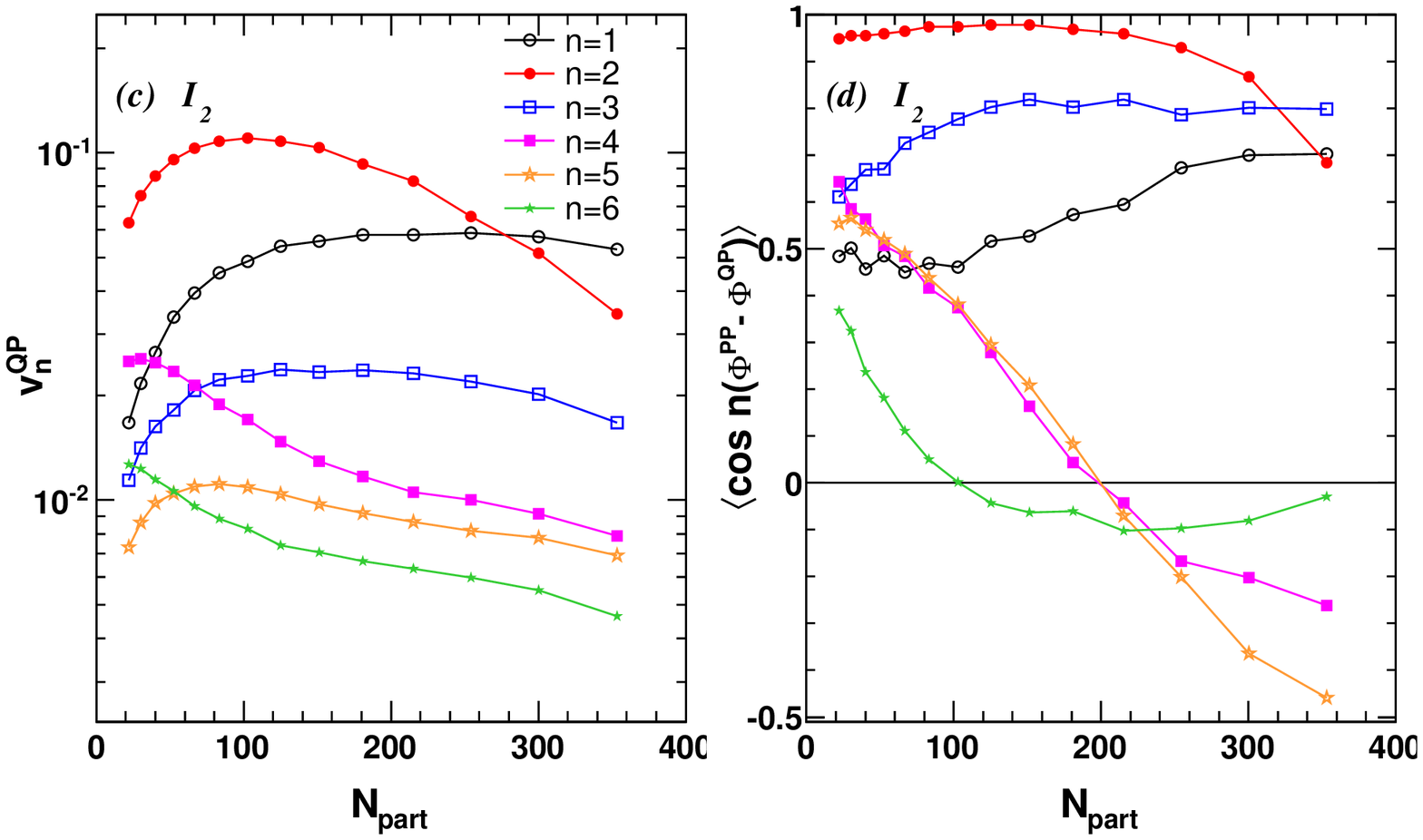}
\caption{(Color online) The centrality dependence of anisotropy coefficients $v_n^{\mathrm{QP}}$ (left panels) and correlation between the participant plane and quenching plane $\langle\cos n(\Phi_n^{\mathrm{PP}}-\Phi_n^{\mathrm{QP}})\rangle$ (right panels) for $I_1$ type of path-length dependence (top panels) and $I_2$ types of path-length dependence (bottom panels). Note that the values of $v_n^{\mathrm{QP}}$ are positive by construction according to Eq.~\ref{eq:4}.}
\label{fig:3}
\end{figure}

Figure~\ref{fig:3} (a) and (c) summarize the centrality dependence of $v_{n}^{\mathrm{QP}}$ for $n=1$--6 and for $I_1$ and $I_2$, respectively. Significant $v_n^{\mathrm{QP}}$ signals are observed for $n\leq3$, while higher-order $v_n^{\mathrm{QP}}$ are usually smaller than 1\%. The $v_2^{\mathrm{QP}}$ and $v_4^{\mathrm{QP}}$--$v_6^{\mathrm{QP}}$ all show strong centrality dependence, while the $v_1^{\mathrm{QP}}$ and $v_3^{\mathrm{QP}}$ show little centrality dependence for $N_{\mathrm{part}}>100$. Interestingly, the value of the $v_1^{\mathrm{QP}}$ is consistently larger than that for $v_3^{\mathrm{QP}}$, and it even exceeds $v_2^{\mathrm{QP}}$ value in most central collisions. This behavior suggest that the path-length dependence of energy loss and initial dipole asymmetry from fluctuations corroborate to produce a large $v_1^{\mathrm{QP}}$. This large $v_1^{\mathrm{QP}}$ is expect to contribute to the high-$\pT$ $v_1$ signal observed by the ATLAS Collaboration~\cite{ATLAS}. Figure~\ref{fig:3} also shows that the $I_2$ type of path-length dependence induces significantly larger $v_n^{\mathrm{QP}}$ than that for $I_1$: the increase is almost a factor of two for $n=1$ and $n=3$. This is also observed in other studies before~\cite{Jia:2010ee,Marquet:2009eq}.

Figure~\ref{fig:3} (b) and (d) summarize the centrality dependence of $\langle\cos n(\Phi_n^{\mathrm{PP}}-\Phi_n^{\mathrm{QP}})\rangle$ for $n=1-6$ and for $I_1$ and $I_2$, respectively. As indicated by Eq.~\ref{eq:7}, this represents the reduction factor of the $v_n$ when it is measured relative to the $\Phi_n^{\mathrm{PP}}$. The reduction is small for $n=2$, except in central collisions where it reaches 15\% for $I_1$ and 30\% for $I_2$. However the reduction is significantly larger for $n=1$ and 3, reaching about 50\% for $n=1$ in mid-central collisions. The $\langle\cos n(\Phi_n^{\mathrm{PP}}-\Phi_n^{\mathrm{QP}})\rangle$ value becomes negative for $n>3$ in central collisions, reflecting an anti-correlation between $\Phi_n^{\mathrm{PP}}$ and $\Phi_n^{\mathrm{QP}}$ (already shown in Fig.~\ref{fig:2}). Interestingly, $\langle\cos n(\Phi_n^{\mathrm{PP}}-\Phi_n^{\mathrm{QP}})\rangle$ values for $n=1$ are always smaller than that for $n=3$ (more misalignment), while $v_1^{\mathrm{QP}}$ is always larger than $v_3^{\mathrm{QP}}$.

The dispersion between the $\Phi_n^{\mathrm{QP}}$ and $\Phi_n^{\mathrm{PP}}$ has important implications on the factorization relation Eq.~\ref{eq:6b}. The factorization of $v_{n,n}$ into $v_n$ is obviously valid for correlations between two low $\pT$ particles (soft-soft correlation) as both are modulated around $\Phi_n^{\mathrm{PP}}$. The factorization should also be valid for correlation between a low-$\pT$ particle and a high-$\pT$ particle (soft-hard correlation) since it involves the projection of the $v_n$ onto $\Phi_n^{\mathrm{PP}}$, i.e. $v_{n,n}(\pT^{\mathrm a},\pT^{\mathrm b}) = v_n(\pT^{\mathrm a})v_n^{\mathrm{QP}}(\pT^{\mathrm b})\langle\cos n(\Phi_n^{\mathrm{PP}}-\Phi_n^{\mathrm{QP}})\rangle= v_n(\pT^{\mathrm a})v_n(\pT^{\mathrm b})$. Experimental data indeed support this~\cite{ATLAS,CMS:2012wg}. However the correlation between two high-$\pT$ particles from two independent hard-scattering processes (hard-hard correlation) is expected to be larger than the product of the two single particle $v_n$:
\begin{eqnarray}
\label{eq:8}
v_{n,n}(\pT^{\mathrm a},\pT^{\mathrm b}) &=& v_n^{\mathrm{QP}}(\pT^{\mathrm a})v_n^{\mathrm{QP}}(\pT^{\mathrm b})\\\nonumber
&=&\frac{v_n(\pT^{\mathrm a})v_n(\pT^{\mathrm b})}{\langle\cos n(\Phi_n^{\mathrm{PP}}-\Phi_n^{\mathrm{QP}})\rangle^2}\;.
\end{eqnarray} 
Therefore, the factorization can not work simultaneously for soft-soft, soft-hard and hard-hard correlations.

\begin{figure}[t]
\includegraphics[width= 1\linewidth]{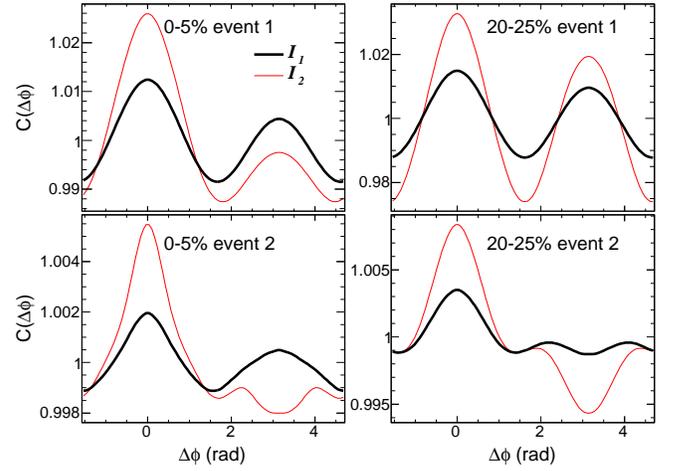}
\caption{(Color online) The expected long-range structures for correlations between two high $\pT$ particles from independent hard-scattering processes. They are shown for two typical events in 0-5\% centrality interval (left panels) and 20-25\% centrality interval (right panels); each of them should be regarded as the distributions obtained for many events with identical initial geometry.}
\label{fig:4}
\end{figure}

The large anisotropy coefficients $v_n^{\mathrm{QP}}$ also has important consequences for the ``ridge'' observed in two-particle correlations~\cite{Abelev:2009qa,ATLAS,CMS:2012wg}. This ``ridge'' is understood to be the result of the constructive contribution of harmonics at $\Delta\phi\sim0$. In the literature, it is referred to as either the ``soft-ridge''~\cite{Daugherity:2008su,Gavin:2008ev} for soft-soft correlation or ``hard-ridge''~\cite{Abelev:2009qa,Moschelli:2009qr} for soft-hard correlation, respectively. Here we show that the correlation between two independently produced high-$\pT$ jets can also produce the ``ridge''-like structure. This ``hard-hard ridge'' can be calculated on a probability basis event-by-event by simply self-convoluting the $R_{\mathrm {AA}}(\phi)$ distribution like Fig.~\ref{fig:1} (e). Examples of these structures are shown in Fig.~\ref{fig:4} for two representative events in both 0-5\% and 20-25\% centrality intervals. The magnitude of the ridge, as well as the away-side shape changes dramatically from event to event. They also changes a lot between the $I_1$ and $I_2$ types of path-length dependence jet absorption. 

\begin{figure}[t]
\includegraphics[width= 0.95\linewidth]{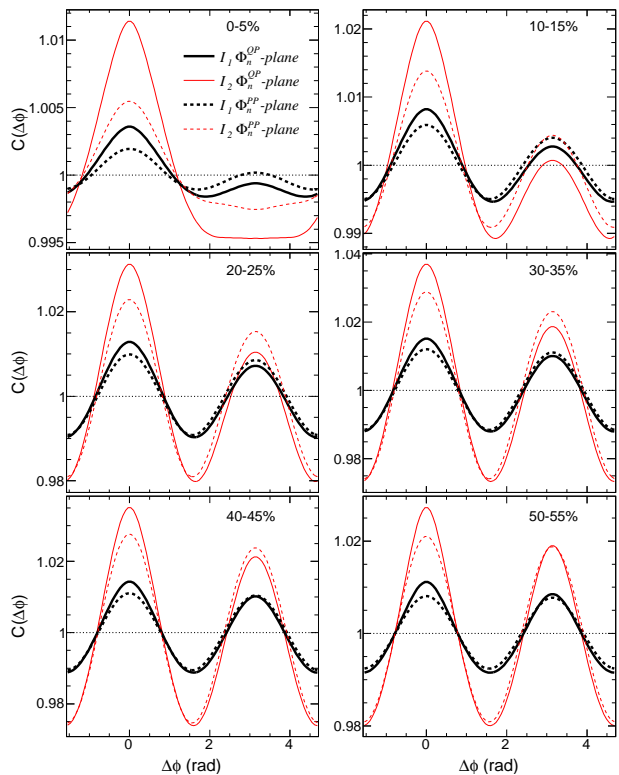}
\caption{(Color online) The expected long-range structures for correlations between two high $\pT$ particles from two independent hard-scattering processes (solid lines) and those calculated from single particle $v_n^{\mathrm{QP}}$ relative to participant planes (dashed lines) for various centrality intervals. They are average distribution over many events for a given centrality intervals. The thick (thin) lines denote the $I_1$ ($I_2$) type of path-length dependence.}
\label{fig:5}
\end{figure}

Figure~\ref{fig:5} show the long-range structures (solid lines) obtained from the jet absorption model, averaged over many events. The ridge magnitude increases with centrality to about 1.5\% (4\%) for $I_1$ ($I_2$) path-length dependence in mid-central collisions. This signal should be measurable with the large statistics dataset from LHC. The dashed lines in Fig.~\ref{fig:5} show the 2PC predicted from the $v_n$ measured relative to $\Phi_n^{\mathrm{PP}}$. Clearly the misalignments between $\Phi_n^{\mathrm{QP}}$ and $\Phi_n^{\mathrm{PP}}$ reduces the ridge magnitude. The reduction is almost 50\% in most central collisions, but decrease to about 20\% in mid-central collisions. This suggests the difference between the measured ridge and those predicted by the event plane method could be large and measurable. 

\section{Conclusion}
The anisotropy of high-$\pT$ particle is studied in a simple jet absorption framework with event by event fluctuating geometry. The harmonic coefficients $v_{n}$ are found to be significant for $n=1-3$ ($>1\%$) but become very small for $n>3$. The correlation between the quenching plane and participant plane are studied. A strong de-correlation is found for $n=2$ in central collisions and for $n=1$ and $3$ over the full centrality range. The correlations become negative for $n>3$ in central collisions. This de-correlation, if also confirmed between the event plane and the quenching plane (e.g via hydrodynamic model that has dijets embeded), is expected to break the global factorization of the two particle Fourier coefficient $v_{n,n}$ into the $v_n$ for the two single particles. It would also imply that the high-$\pT$ $v_n$ measured relative to the event plane could be significantly smaller than the true anisotropy from path-length dependent jet energy loss. These jet quenching $v_n$ also give rise to long range ``ridge'' structure in two-particle correlations. The predicted ridge amplitude is on the order of 0.5-4\% depending on the centrality and functional form of the $l$ dependence of the energy loss, and should be measurable at the LHC using the correlations between two high-$\pT$ particles with a large rapidity separation. Our study bear some similarities to Ref.~\cite{Zhang:2012mi}. However, Ref.~\cite{Zhang:2012mi} uses a cumulant expansion framework instead of Monte Carlo Glauber model for initial geometry, and that it focus on the soft-hard ridge instead of the hard-hard ridge in our case. 

Discussions with Jinfeng Liao is acknowledged. This research is supported by NSF under award number PHY-1019387.
 % do not change
\end{document}